# eXtreme Multiplex Spectrograph: An efficient mechanical design for high-demanding requirements

S. Becerril[1*], K. Meisenheimer[3], C.M. Dubbeldam[2], R. Content[2], R.R. Rohloff[3], F. Prada[1], T. Shanks[2], R. Sharples[2]

[1] Instituto de Astrofísica de Andalucía (IAA-CSIC), Glorieta de la Astronomía s/n, Granada, Spain

[2] University of Durham, Durham, England

[3] Max-Planck-Institut für Astronomie (MPIA), Heidelberg, Germany

## ABSTRACT

XMS is a multi-channel wide-field spectrograph designed for the prime focus of the 3.5m Calar-Alto telescope. The instrument is composed by four quadrants, each of which contains a spectrograph channel. An innovative mechanical design –at concept/preliminary stage- has been implemented to: 1) Minimize the separation between the channels to achieve maximal filling factor; 2) Cope with the very constraining space and mass overall requirements; 3) Achieve very tight alignment tolerances; 4) Provide lens self-centering under large temperature excursions; 5) Provide masks including 4000 slits (edges thinner than 100µ). An overview of this extremely challenging mechanical design is here presented.

* Contact S. Becerril (santiago@iaa.es) for further information.

## 1. INTRODUCTION AND SCOPE

This paper presents an overview of the mechanical design of XMS, a multi-channel wide-field spectrograph for the prime focus of 3.5m Calar Alto telescope. The instrument segments the unvignetted 1-degree field available into four equal quadrants, each of which contains an identical spectrograph channel. Each mask covers all four channels giving spectra for up to 4000 objects in a single exposure.

An innovative mechanical design approach has been implemented to cope with the tight constraints in terms of space envelope, minimization of the separation between channels, overall mass and position tolerance over the whole temperature range at working conditions. And, last but not least, the masks are highly demanding in terms of design and manufacturing since each of them must include up to 4000 slits on a 100µ-wide sheet, keeping the shape errors lower than 60µ.

## 2. MECHANICAL LAYOUT

The XMS instrument can be broken down into the following mechanical subsystems (see Figure 1 (left)).

- **Mask Exchange Unit** (MEU) (**): It changes the masks according to the astronomer's needs. It interfaces with the K3 prime focus corrector and includes the Acquisition & Guiding Unit.

- **Collimation Main Stack** (CoMS): It includes lenses from L1 to L5
- **Shutter Unit** (**): It works as both an instrument shutter and a Hartmann Shutter.
- **Manual Gratings Unit** (MGU): It includes the prisms, as well as the last lens (L6) from the collimation stage and the first lens (L7) from the camera stage.
- **Camera Focus Unit** (CFU): This mechanism allows optical fine focusing of each XMS spectrograph channel independently by tuning Z position of doublet L8 & L9.
- **Camera Main Stack** (CaMS): It includes the lenses from L10 to L13.
- **Detector Dewar** (**): This subsystem provides appropriate conditions (170K, vacuum) for optimal operation of the four CCDs (one per channel).
- **Detectors Mount**: The detector mount allows tip-tilt alignment of the detectors during AIV and is enclosed within vacuum environment inside the Detector Dewar.
- **Optical Bench** (OB) (**): It is the main mechanical structure supporting all the XMS optomechanical components (except for the MEU).

The present work focuses on the most critical subsystems. The CaMS, the CFU and the Detectors Mount have been developed at interfaces-and-envelope level. Regarding the optomechanics, the CoMS have been selected as the most critical according to the requirements and constraints applicable to. Since the feasibility of the CoMS is demonstrated, the rest of optomechanics shall be developed at the Preliminary Design stage. Each of the subsystems marked with (**) from the list above is one common item for the 4 channels. For the rest, there are 4 items (one per channel).

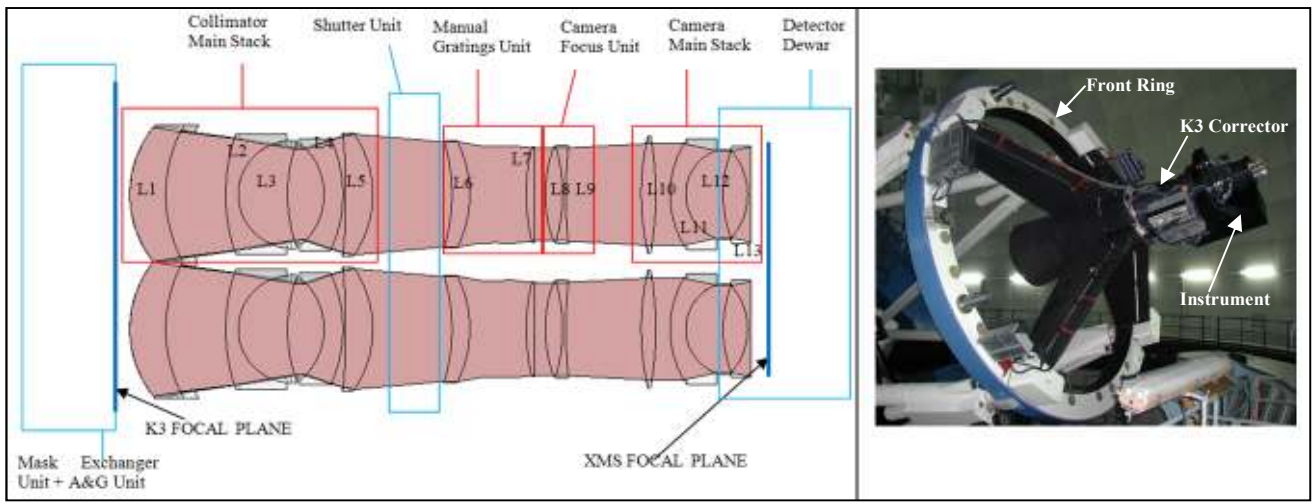

Figure 1. Left: Mechanical layout of the instrument. Right: The 3.5@Calar-Alto Prime Focus station.

## 3. MECHANICAL REQUIREMENTS AND BOUNDARY CONDITIONS

Within the present work, the typical coordinate system used presents the axis Z parallel to the optical axis; while axes X and Y define a plane orthogonal to the Z-axis. Next, the mechanical requirements are listed (Table 1).

Table 1. Mechanical requirements and boundary conditions concerning the XMS instrument.

| Requirement | Value | Comment |
|---|---|---|
| Location | Prime Focus | Any orientation of the instrument as regard the gravity vector is possible (except for the instrument looking downwards) |
| Mechanical Interface | K3 Corrector Flange | K3 Corrector is a 3-lenses barrel providing the Focal Plane for some instruments at Calar Alto. |
| Max Overall Mass | 250 kg on K3 Corrector Flange | |
| Max Overall Mass | 250 kg on the Front Ring (Figure 1 (right)) | Evenly distributed |
| Z envelope | 750mm | Risk of collision against the dome, if exceeded. |
| Residual alignment tolerances | see Table 2 | |
| Thermal-Optical Analysis inputs | Small effects on the image quality over the working temperatures range | Aluminum for the optomechanical mounts and a material with a CTE of 13 x 10-6/°C for the OB. |
| Mean Working Temperature | 6.3ºC | |
| Working Temperature Range | From -15ºC to 20ºC | |
| Maximal Temperature Variation | 5ºC in 3 hours | |
| Maximization of sub-fields of view (s-FoV) | Inner cross-shaped gap between L1 lenses must be minimized (Figure 3) | Each of the spectrograph channels gets ¼ of the FoV. A cross-shaped gap is produced between the 4 L1 lenses. |
| XY envelope diameter | < 820mm | Otherwise, some obscuration of the beam is produced. |
| Local Dome seeing | Thermal distortion to be avoided | Water-cooled control electronics |
| Pressure range | From 700 to 1050 hPa | |
| Minimal operation lifetime | 10 years | |

Table 2. Position tolerances of the optics after alignment. "Grism Boxes" refers to MGU, while "Aspheric lens" refers to L5 lens.

| Type of error | Element | Tolerances | |
|---|---|---|---|
| Resulting PSF 50% EED for 98% confidence | | 0.20" (goal) | 0.34" |
| De-center XY | Lenses | 15μm | 30μm |
| De-center XY | Grisms Boxes | 15μm | 30μm |
| De-focus Z | Aspheric Lens | 10 μm | 20 μm |
| De-focus Z | Other lenses | 15 μm | 30 μm |
| De-focus Z | Detector | 10 μm | 20 μm |
| De-focus Z | Grisms Boxes | 15 μm | 30 μm |
| Tip and tilt | Lenses | 0.015[a] | 0.030º |
| Tip and tilt | Grisms Boxes | 0.007º | 0.014º |

# 4. XMS MECHANICAL DESIGN

## 4.1. General issues

The combination of requirements presented above sets a **demanding scenario** for the mechanics of the instrument. The mass and envelope requirements lead to driver guidelines (applicable to every subsystem) consisting of high compactness and lightweight. Concerning the OB and the MEU, high stiffness is also a very important issue. The Figure 2 shows some general views of the whole instrument.

Regarding the position tolerances, **self-centering radial mounts** are, in practice, necessary. Otherwise, the lenses would get de-centred between each other more than specified. The position tolerances (after-alignment) are so tight that manufacturing cannot ensure them. It has led to the optomechanics design to include three re-machinable parts for each lens. Therefore, there shall be a trial/error process for the lenses to be well aligned inside the subsystem in which they are housed.

Furthermore, each channel will include dedicated subsystems (CoMS, CaMS, etc…). Thus, it shall be independently aligned and integrated with respect to the other channels. The current approach allows each channel's optomechanics to be divided into different subsystems, which can be handled and mounted at subsystem level. Once this AIV stage has been successfully completed, the subsystem must be integrated in the system by being mounted to the Optical Bench and aligned with respect to a reference. An alternative approach where the optomechanics was transversally implemented would lead to physical "frames" (or optomechanics housings) shared by the four channels, which would probably require the full assembly and alignment of the lenses there involved prior to start with the integration of the following optomechanical subsystem. This would have important drawbacks in terms of AIV working planning and schedule: 1) work in parallel would be very much handicapped; 2) complex division of the workpackages (for eventual external subcontracts); 3) longer AIV periods; 4) no distinction between AIV at subsystem level and AIV at system level; 5) Less robust AIV schedules to eventual delays upon the packages externally subcontracted.

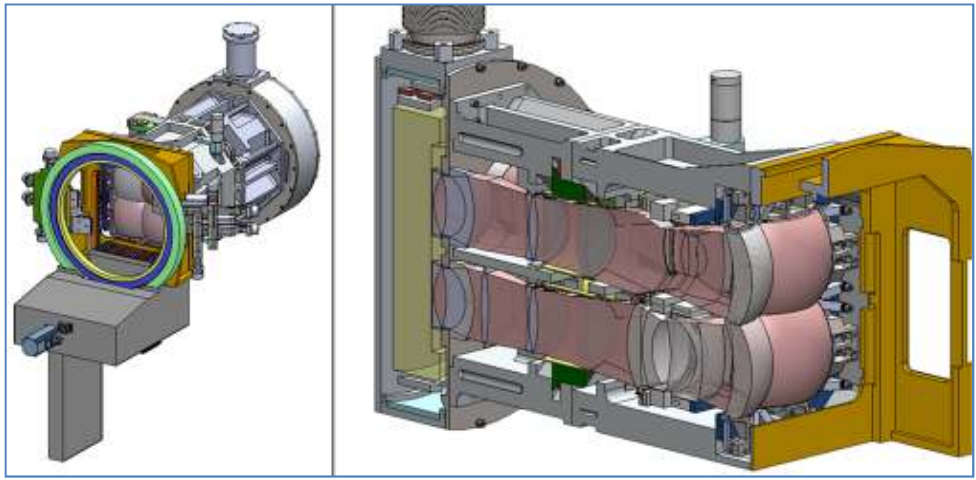

Figure 2. Left: General view of the entire instrument. Right: Longitudinal cut-view.

## 4.2. Optomechanics Description

The CoMS includes the largest lenses of the optical layout. Therefore, space constraints are also here the most severe. Moreover, that subsystem presents the highest amount of lenses amongst all the optomechanical stacks.

The FoV considered for the XMS instrument is squared (see Figure 3), which leads to rather square-shaped lenses, as well as non-axis-symmetric geometry. In addition, the effective beam section is squared at the mask and it remains still sharply squared at the L1 Lenses. As long as the beam comes forwards within the collimation stage of the optics, the beam section becomes rounder on the corners and, in general, smaller. On the pupil, the beam section is completely round and, from this point on, it gets more and more squared to, finally, become completely square again on the detector (see Figure 4).

All the lenses, except for L1 lenses, have the four areas shown in Figure 4 (right) available for optics holding purposes. In addition, those areas are in opposition, which allow radial preloads to be properly balanced. The design concept applied for all the lenses of the CoMS (except for the L1) is based on:

- A compliant cell where the lens is bonded to. The cell includes 4 radial flexures machined in the bulk (see Figure 5 (detail)). This concept allows the lens to remain centred as regard to its cell even under temperature variations. The bond also provides axial holding. This solution provides high compactness, few mechanical parts and low mass. Obviously, the lens position cannot be changed within the cell for alignment. It is, in fact, the cell which is tuned in position (XY de-center).

- Three re-machinable parts for alignment purposes provide easy adjustment on tilt (over X and Y axis) by means of proper re-machining. As a reminder, this compensation is necessary for alignment since the tolerances mentioned above cannot be reached directly by manufacturing.

- Alignment elements not integrated in the optomechanics: The current approach consists of an alignment kit (not yet designed) that can be coupled to each optomechanical mount to tune the XY position (de-centre) of a certain lens cell. Once the lens cell is well positioned, it is fixed to its interface and the alignment kit removed.

Regarding the optomechanics attached to the Detector Dewar, mention has to be done of the fact that the L13 lenses, even if being the last ones of the optical layout, cannot work as dewar windows since they are only 3mm thick on their thinner section. Indeed, this does not guarantee the integrity of the L13 lenses under the pressure breakage. Alternatively, there is the doublet L11 & L12, which presents better features for such purpose. In particular, L11 has been chosen for being the dewar window.

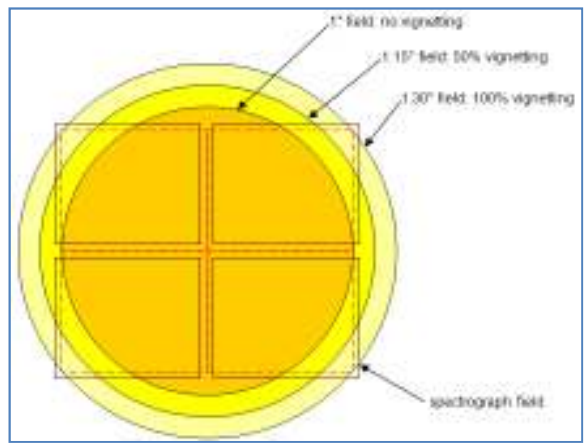

Figure 3. XMS field with inner cross-shaped gaps.

### 4.2.1. Collimator Main Stack

The Figure 5 shows a general view of the CoMS. Its design is based on a stack of cells mounted on each other, except for the L1 Mount. The latter, due to its special features, is treated in a different way. The CoMS Main Frame is the part interfacing with the OB. This stack-based concept is very compact related to its XY envelope. Moreover, this solution is quite natural for the present optics since the sizes of lenses L2, L4 and L5 are quite similar. These lenses are separated from each other by means of three small spacers.

Each lens cell includes three bronze pads, which, in fact, interface to the spacers of the mount previously integrated. For instance, the L2 Mount provides the spacers on which the bronze pads of the L4 Mount are in contact. Once the position of L4 is determined, it is fixed by means of three screws which are aligned with the bronze pads. Thus, no deflection is induced on the lens cell. All the pads can be dismounted and re-machined in order to provide to the lens appropriate tilt and Z-position.

As mentioned before, each cell is bonded to the lens through 4 radial flexures (see Figure 5 (detail)), which provide self-centering performance to the lens mount.

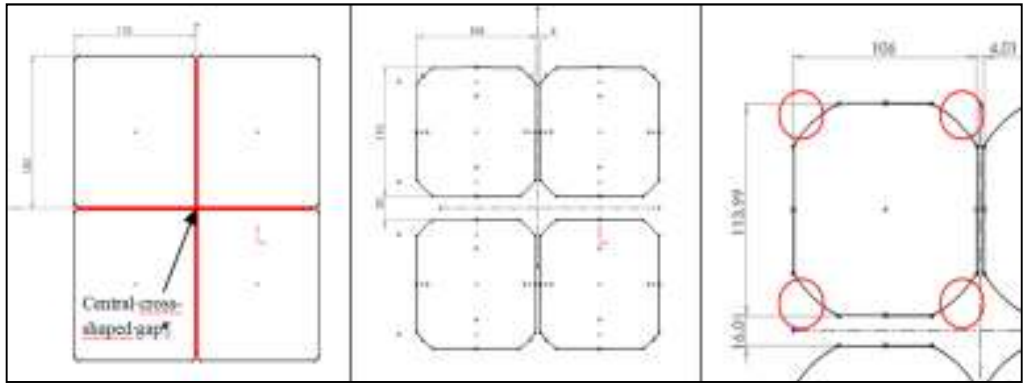

Figure 4. Left: Clear aperture on first optical surface of L1 lenses. Middle: Clear aperture on first optical surface of L2 lenses. Right: Example of areas (red) available for lens holding (L5 lens).

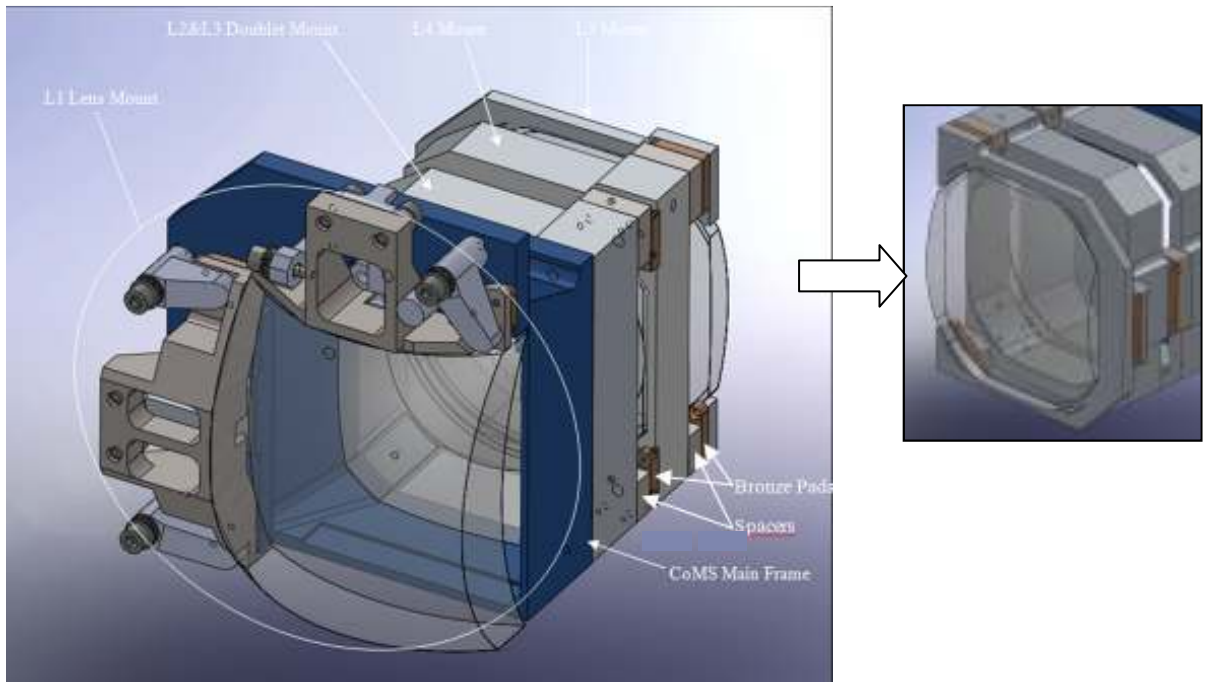

Figure 5. General view of the CoMS. In the detail: the L5 lens cell, with 4 flexures for self-centering. Note that, for each lens, two pads (and spacers), very close to the cross-shaped gap between channels.

### 4.2.2. Entrance Lens (L1) Mount

L1 lenses need a different optomechanical approach since the areas shown in Figure 4 (right) are not available. However, the self-centring feature must be respected,. Additionally, independent optomechanics for each L1 lens has been stated as a design guideline. The Figure 6 (left) shows a general view of the L1 Lens Mount integrated on the CoMS Main Frame.

The lens is held only through the outer lateral faces, thus the inner cross-shaped gap being minimized. Some metal parts must be bonded to the outer lateral faces of the lens in order to provide appropriate interfaces for support. Thus, no extra machining features will be implemented on the glass, which will prevent it from getting overstressed during the manufacturing process. Such metal parts shall have a CTE close to that of the L1 lens. In practice, it means that such metal parts shall be made of Ti6Al4V.

The self-centering feature (see Figure 6 (right)) is achieved by means of a couple of aluminum beams that, together with the Ti6Al4V parts, play the role of two bimetallic compensators (Ti6Al4V/Aluminum). One end of the beams is attached to the CoMS Main Frame, while the other end is attached to the respective Ti6Al4V bonded part. The length of the beams is such that any change in length of the glass (plus the Ti6Al4V part) is compensated by the change in length of the aluminum beam.

Since the compensators will work along orthogonal directions each other, they would induce high stresses on the lens, if they were rigid beams. That's why each of them includes a joint at each end of the beam.

The Figure 7 (left) shows how this compliant configuration allows the L1 lens to remain centred while no overstress is induced on the glass. The solution here presented for self-centering (Figure 7 (right)) does not constrain the rotation over Z-axis. The latter is constraint by an anti-rotation stop.

The constraint of translation along Z and rotations over X and Y is made by means of three bronze pads which contact on the CoMS Main Frame. Note that the joints of the aluminum beam are also compliant so that the contact of the bronze pads against the CoMS Main Frame is not prevented. The L1 Lens mount must be preloaded against the anti-rotation stop and against the CoMS Main Frame (through the bronze pads) at any orientation with respect to the gravity vector. Belleville washer stacks are used to provide the appropriate preloads. Preload forces have to be kept as small as possible whenever key mechanical contacts are ensured. The static constraints analysis has shown that preload can be as low as 20N (safety factor: 2). Because of this low preload, the L1 lens cell is able to slide on the CoMS Main Frame.

Finally, the cross-shaped gap can be roughly as small as 2mm. This would include all the position and manufacturing errors applicable. At the Preliminary Design stage, a complete error budget should be done. Then, the width of such gap will be defined more accurately.

### 4.3. Detector Dewar

The very tight constraints above mentioned means that only 150mm are left from the CCD sensitive surface up to the rearmost part of the Detector Dewar in order to avoid any collision with the dome. The preferred approach is based on a Cryotiger cooler. As compared with LN2-bath cryostats, it allows higher compactness and fewer parts, thus the resulting mass being appreciably lower. Additionally, noticeable cost saving in terms of manpower is also provided since no daily maintenance is required. Maintenance is limited to some tasks to be done on quarterly basis. As main drawbacks, some issues concerning the coolant lines from the compressor to the cryocooler have to be solved, as well as the location of the compressor. These issues shall be thoroughly studied at Preliminary Design stage.

The optomechanics of lenses from L10 to L13 interfaces with the Front Cover. Each channel will include a dedicated, independent stack of lenses L10 to L12. From the doublets L11&L12, the L11 lenses work as dewar windows, while L12 lenses are directly glued to the respective L11. Therefore, the four L13 lenses shall be entirely integrated in the vacuum environment.

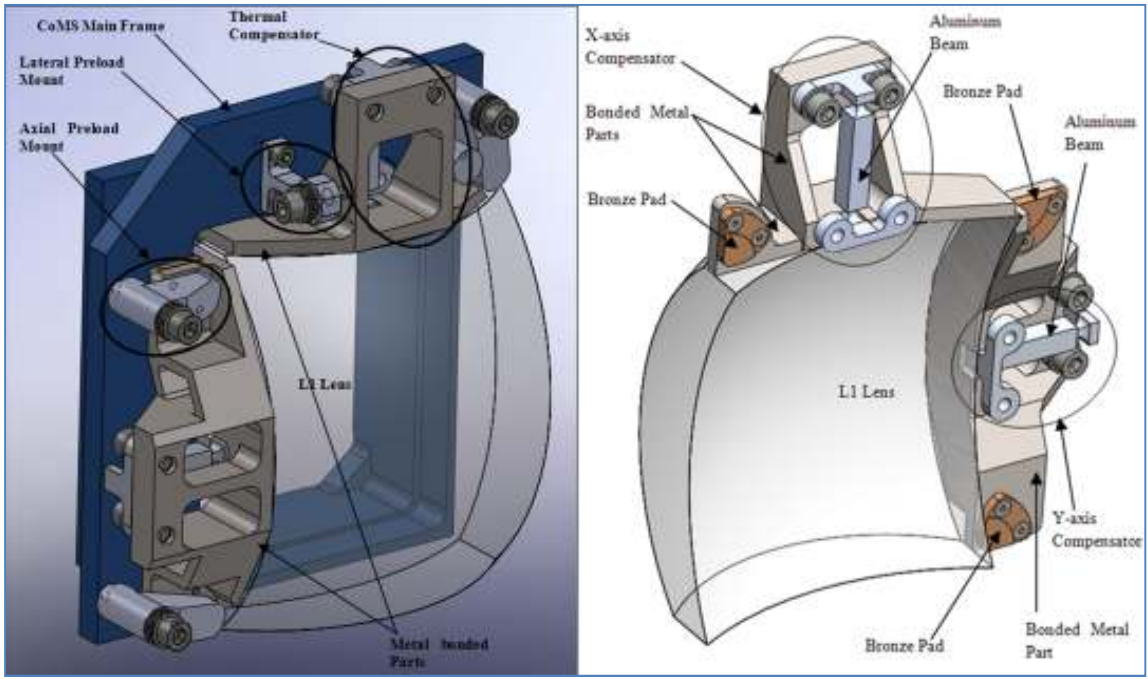

Figure 6. Left: General view of the L1 lens mount integrated on the CoMS Main Frame. Right: L1 lens compensators for self-centering issues.

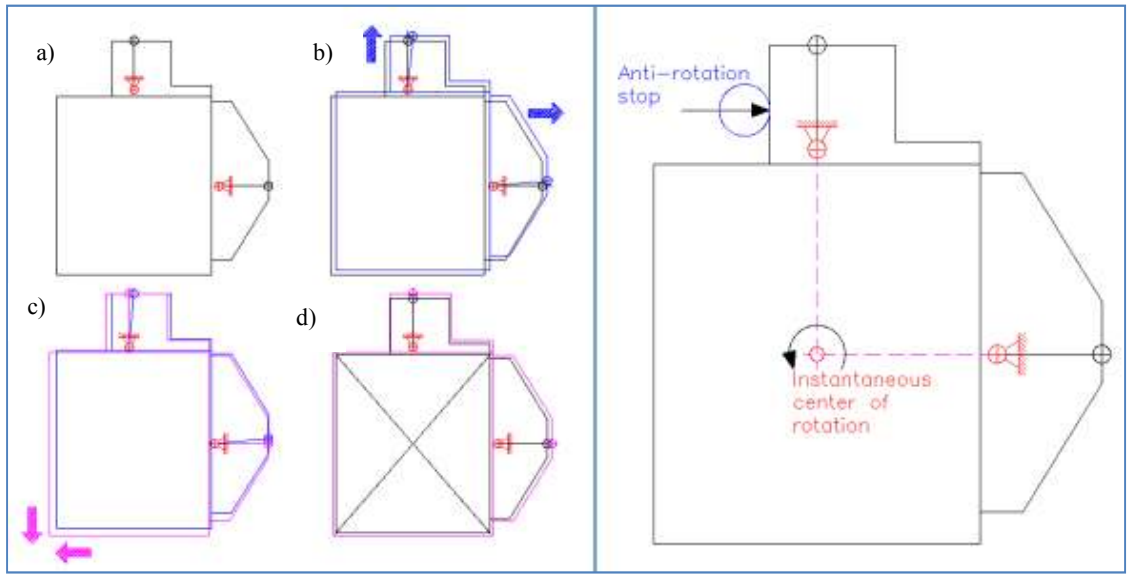

Figure 7. Left: A sequence about how the L1 mount keeps self-centered under a temperature variation. a) start position; b) expansion of the beams produce a rotation over the beam joints (overstresses avoided); c) expansion of the glass and bonded parts; d) final position as compared with start position (centering is preserved). Right: DoFs constraint on plane XY.

As well as the L11 lenses, the Detector Mount shall interface with the Front Cover (inner surface). This way, the relative position between the lenses and the CCDs is much better controlled. That's why deflections of the Front Cover must be minimized, which implies it must be made of Stainless Steel. Thus, the rest of the Vacuum Vessel parts can be made of aluminum, which leads to a noticeable mass save.

According the analyses implemented, a Cryotiger device PT30 fits well the CCDs needs (170K). Concerning the structural behavior of the Vacuum Vessel under vacuum, the FEA analyses show that the center of the L11 lenses move by 57μ and tilt by 0.028º (which is within tolerances (0.030º)). Some kind of compensation shall be anyway implemented during the AIV phase in order to keep larger margins for the position tolerances. The maximal stresses produced by the pressure breakage are small (35MPa for the Front Cover and 42MPa for the rest); therefore, the materials involved (stainless steel and aluminum) will be by far below the yield limit.

The Cryotiger-based concept (Figure 8) consists of a cylinder-shaped vacuum vessel that houses the Detectors Mount. Two covers provide appropriate sealing for vacuum preservation, as well enabling access for maintenance and integration purposes. For envelope reasons, both the cryocooler (by means of an adapter) and an electropneumatic valve are attached to the Vacuum Vessel Body. The electropneumatic valve allows removing the pumping system once the vacuum level is achieved. A gauge measures the vacuum level and feedbacks the vacuum control unit during the pumping protocol.

The Cryotiger cold tip is rigidly attached to a Thermal Sink made of aluminum, which, in turn, is thermally attached to the Detectors Mount by means of flexible copper straps (not shown on figures). The Thermal Sink provides an interface with low thermal gradient. Otherwise, due to the cryocooler location, noticeable thermal gradients would be transferred to the Detectors Mount, which are not desirable from control point of view. The copper straps –annealed treatment proposed- allow further smoothing of the thermal gradient on the CCDs Mount, as well as the necessary compliance to avoid overstresses due to thermal relative displacements.

The Thermal Sink is attached to the Front Cover by means of two bi-metallic support pads (not shown on figures), which include parts made of G10 in order to minimize the conduction heat losses. By appropriate design of the bi-metallic feature (material choice, length of the components) the Thermal Sink axial position keeps invariable. The location of the two support pads are aligned and centred on the symmetry plane of the Cryotiger. Thus, those supports do not need to be compliant along the Y direction.

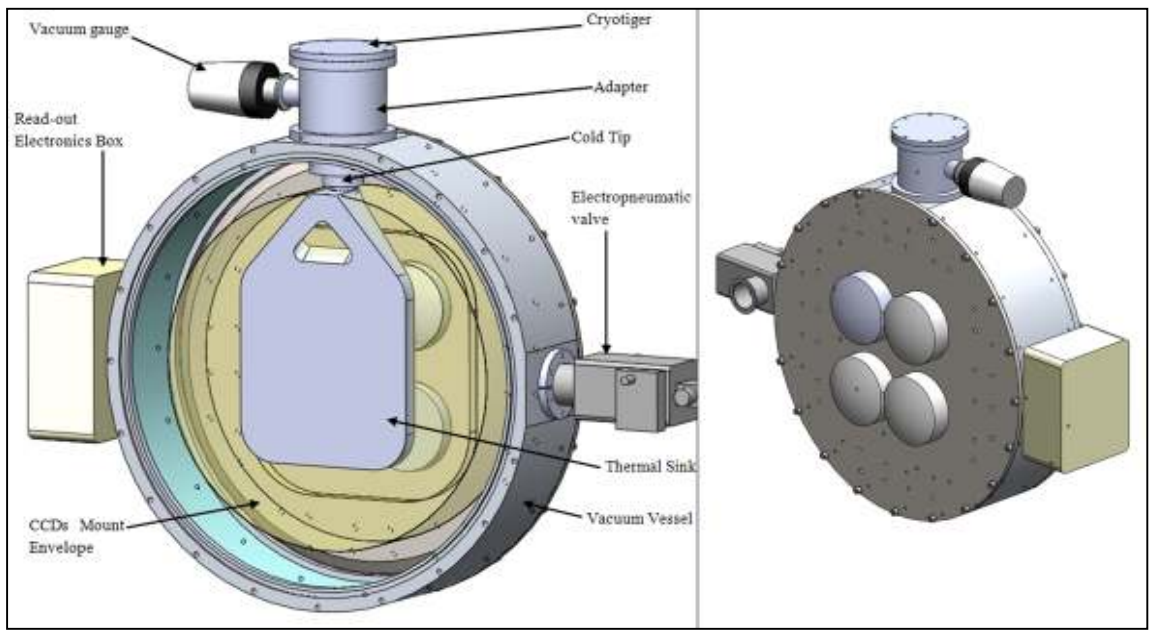

Figure 8. Left: General view of the Cryotiger-based dewar inside (Rear Cover removed). Right: General view of the Cryotiger-based dewar (L11 optomechanics must interface the Front Cover).

### 4.3.1. LN2-bath dewar alternative

This approach can only fit the tight space constraints if the LN2 Tank is placed around the Detectors Mount. The dewar gets noticeably wider, thus the radiation thermal losses being increased with respect to the Cryotiger-based dewar. That's why one layer of radiation shielding is necessary to reach hold-time longer than 24 hours (only ½ LN2 tank can be filled up). In terms of architecture, the vacuum vessel is similar to the one of the Cryotiger-based solution, thus the main difference lying on the overall dimensions. This alternative presents an overload of 25 kg (overall mass requirement not fulfilled) as compared with the Cryotiger-based option but fits, in principle, better the Calar-Alto standards. That's why this has not been completely discarded yet, although the Cryotiger-based option is the baseline here stated.

Figure 9 shows a general cut-view of the Detector Dewar. The Vacuum Vessel is composed by the Front Cover, the Main Cylindrical Body and the Main Rear Cover. An auxiliary cover on the Main Rear Cover (not shown in the figures) is necessary for access and integration of the LN2 Tank and the Rear Radiation Shielding. The LN2 Tank is composed by four parts welded to each other. The rear plate is called the Cold Plate and interfaces with the Detectors Mount by means of thermal linkage (not shown in the figures) composed by several copper straps and a rigid copper part.

Nitrogen vapour originated during the cooldown is evacuated through the Exhaust Vapour Exit Mount (Figure 10). Both in the Exhaust Vapour Exit Mount and the Filling Tube, bellows are foreseen in order to absorb differential displacements without overstressing any of the parts. The Exhaust Vapour Exit Mount must be kept open during operation so that the nitrogen vapour does not produce an excessive overpressure. On the other side, due to the tank geometry, the exit of vapours from the tank through the central Exhaust Vapour Exit Mount must be promoted. This is done by providing 4 exit ducts that are machined within the Cold Plate bulk –no additional welds are required- connecting with the exit pipe. The Filling Tube has not been implemented coaxially to the exit pipe since this may make the vapours exit more difficult. Once the tank is filled up, the Filling Tube shall be blocked by means of a cap.

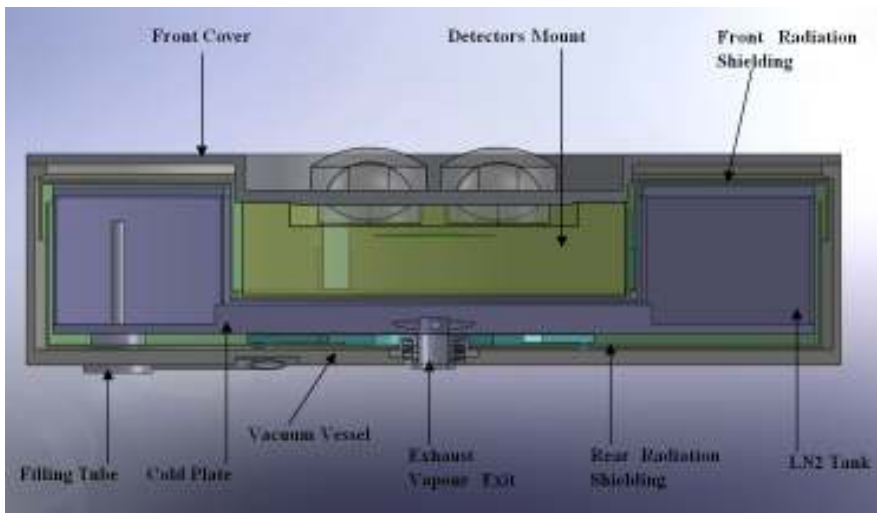

Figure 9: General cut view of the Detector Dewar.

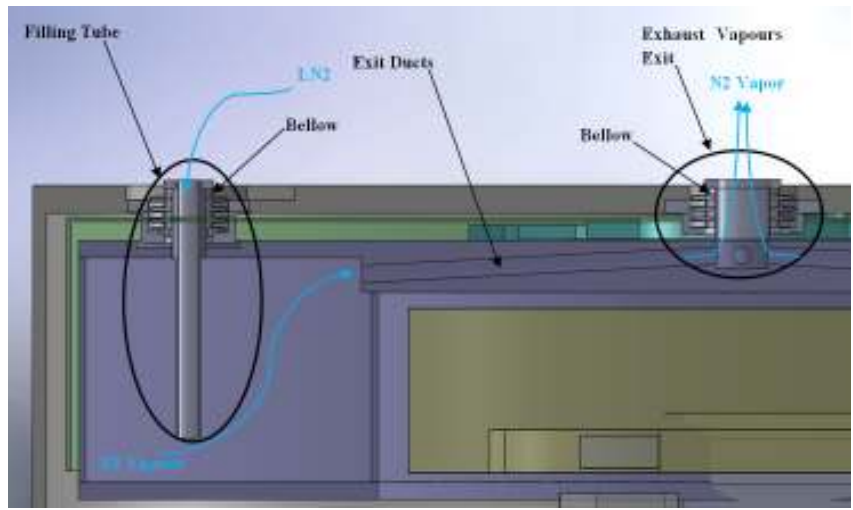

Figure 10: Detail cut view of the LN2 Tank, the Filling Tube and the Exhaust Vapours Exit. Exit ducts machined in the Cold Plate bulk are shown.

**4.4. Optical Bench**

The requirements applicable lead to very light and stiff OB concepts. Moreover, it must fit a CTE around 13x10e-6. The latter condition may be fulfilled by steel alloys but, on the other hand, the instrument mass would be by far off the requirements. Alternatively, Al/Si alloys combine well the requirements above mentioned and it has been chosen as baseline at the present stage.

The design here presented (Figure 11) is based on two parts (bolted to each other) because machining procedures are preferable rather than any other including welding techniques. Indeed, machining procedures induce lower stresses during manufacturing. Furthermore, these stresses can be relieved by appropriate thermal treatments. Stiffening ribs have been implemented wherever there is space available. Especially important are the internal cross-ribbed reinforcements.

Although not all the subsystems have been designed at the present stage, their interfaces with the OB have been already considered. This concept defines some issues concerning the assembly of the different subsystems on the OB:

- Each CoMS shall be mounted on the OB without the MEU on place.
- Both the MGUs and the Shutter Unit can be independently mounted and dismounted with respect to the other subsystems.
- The CFU units must be mounted through rear access to the OB, with the Detector Dewar not mounted yet.

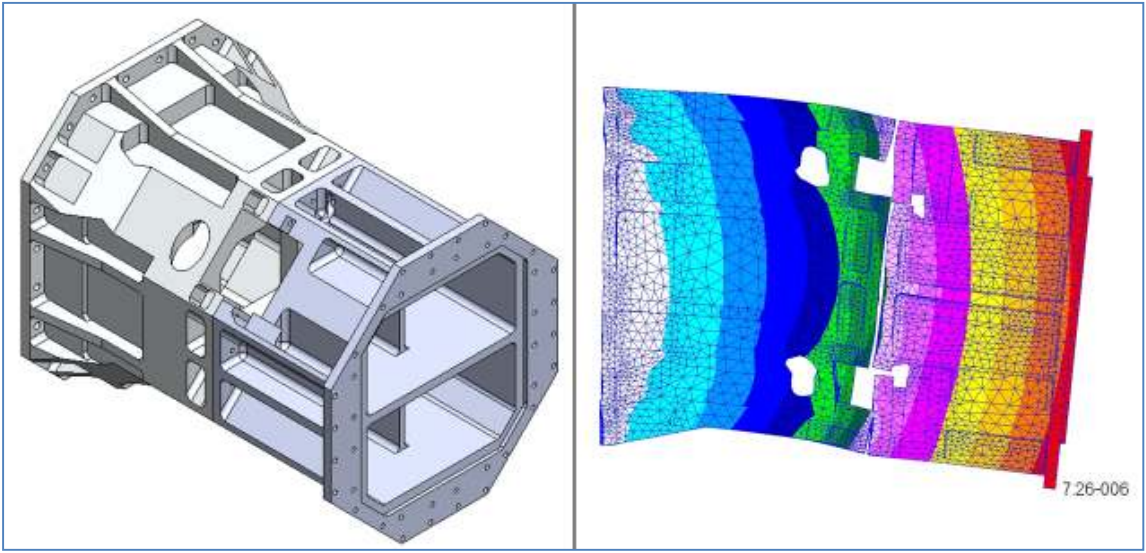

Figure 11. Left: General view of the Optical Bench. Right: FEA Flexure analysis of the OB (unit: meter).

The Finite Elements Analysis (FEA) applied to the OB is necessary for optimisation reasons. The bolted joint of both halves of the OB has been also simulated, thus more realistic results being obtained. In spite of the lightweight design produced, the relative displacements and tilts of the subsystems as regard to each other are limited within tolerances. The maximal flexure of the OB in the worst case is 7.3µm, which appears on the free end (right end in Figure 11 (right)).

Nevertheless, particularly interesting are the values just at the centre of the beam pass-through holes since they are applicable to the centre of the respective lenses attached (L11&L12 Doublet and L13 lenses). Such values are 6.9µm (X direction), 0.5µm (Z direction) and 0.0006º (tilt over Y). The latter two are negligible values while the first one is well within tolerances.

### 4.5. Manual Grating Exchange Unit

The MGU is a unit which includes L6 and L7 as well as the grism. Each MGU must be easily mountable by hand on the OB, being the assembly time and the repeatability on the working position the most important parameters. Although the latter has not been defined yet, the present concept design is proposed on the basis that an isostatic mount is necessary, which avoids any overconstraint on the position.

The mounting trajectory for each MGU will be radial through a dovetail-shaped slot machined on the OB (Figure 12). Once in the slot, the position is defined by the contact items listed below:

- 3 contact items constraining to the plane XY (Figure 12 (right)): These are machined on the dovetail-shaped flange of the MGU (green part).
- 2 contact items constraining both one translation on the plane XY (orthogonal to the slot) and the rotation over Z-axis (Figure 12 (right)). These contact items are machined on one of the two wedge planes of the dovetail-shaped flange.
- 1 contact item (Contact Item 3 in Figure 13 (left)) constraining the translation along the slot. Indeed, the reference part (red colour) prevents the MGU translation along the slot direction.

Fasteners that fix the position of the MGU are listed below (Figure 13):

- Fast Clamp: This easy-to clamp fastener secures the axial position of the MGU along the slot. It is accessible from outside and is fixed by hand.

- Wedge clamp: This wedge accommodates its position to secure the contacts of the dovetail-shaped flange of the MGU. This clamp is tightened by a screw accessible from outside. Some holes have been specifically included in the design of the OB in order to access to the screws of the wedge clamps.

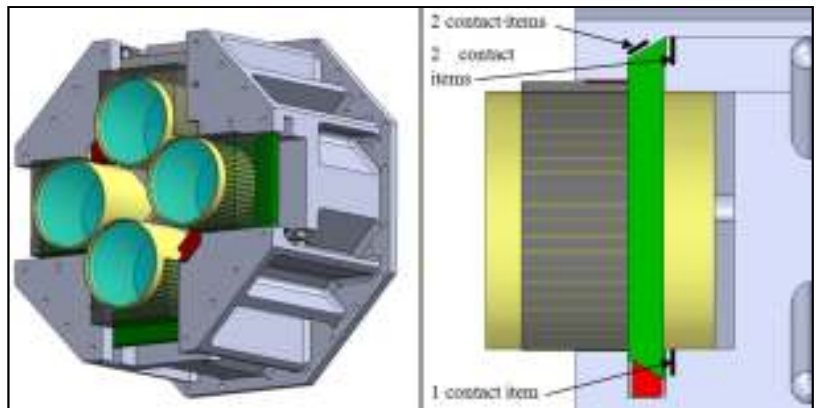

Figure 12. Left: General views of the 4 MGUs inside the OB. Right: Cut side view of one MGU inside its slot. Five of the 6 constraints for isostaticism are shown.

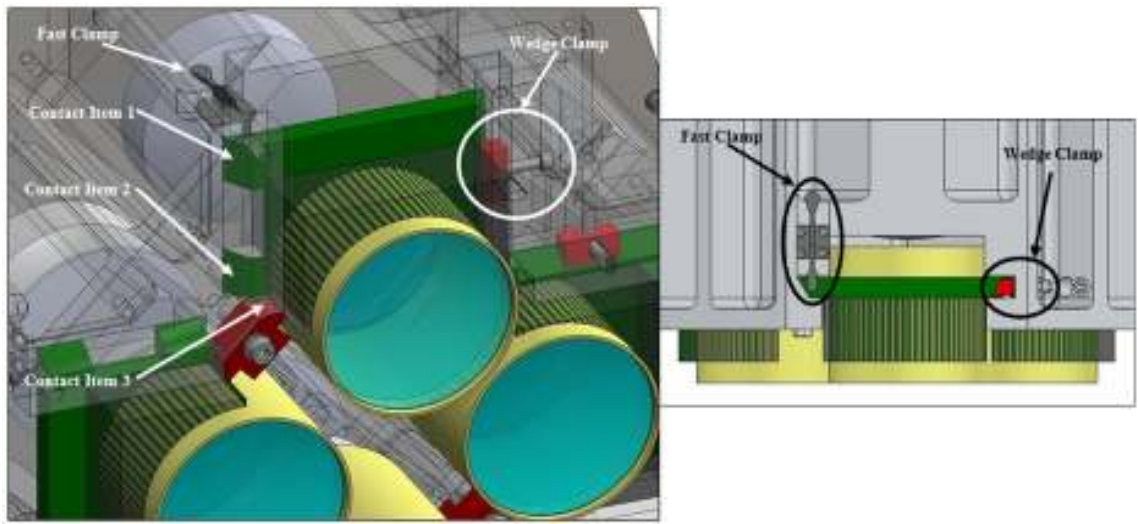

Figure 13. Left: General view showing all the hardware constraining a MGU position, as well as three contact items. Note that Contacts Items 1 and 2 correspond to the 2 contact items on the wedge plane shown in Figure 12. Right: An auxiliary top view for better understanding of the Fast Clamp and the Wedge Clamp.

### 4.6. Shutter Unit

The particular requirements applicable to this subsystem concern the opening time (10s; 1s(target)) and the variation of effective shutter open time across the field (less than 1% , 0.1% (target)). In addition, the Shutter Unit must work also as a Hartmann Shutter.

This concept is based on a single fabric-made frame which goes from one side to the other side of the FoV. Indeed, such concept makes that the pixels being the first to be exposed are also the first to be obscured. The only location available for the Shutter Unit is between L5 and L6, just before the MGUs (60mm far from the pupil, where the beam is almost collimated).

The reinforced fabric-made frame can be rolled on any of two drums. One of the drums is attached to the motor while the other is spring-preloaded in order to keep the fabric tight.

**4.7. Mask Exchange Unit and Acquisition & Guiding Unit**

The MEU contains the mask exchange mechanism and a cassette system to hold spare masks. This mechanism changes the masks depending on astronomer's observations. Each mask shall cover the whole FoV of all four channels. It also contains the instrument's Acquisition & Guiding system (Figure 15).

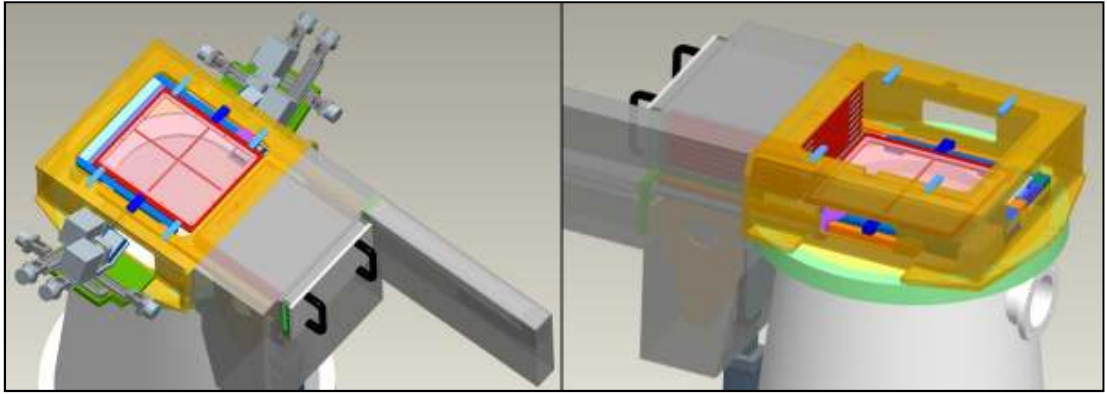

Figure 15. Left: Solid model of MEU with the 4 A&G units included. Each A&G unit can be moved in two directions: in/out and along the reference surface on the mask frame (not visible).The mask is in observing position. Right: MEU is coupled to the Corrector K3 by a rotating flange (green). The mask is extracted from the magazine (not in observing position).

The mask exchange mechanism contains a magazine which holds 8 masks. The masks are loaded into the magazine during daytime. The exchange process consists of two movements: A sideways shift moves the mask out of the magazine. Then a move along the optical axis brings the mask to a fixed position in the focal plane of the telescope. In the final position the defining surface (edge of slitlets) must not deviate more than +/-60 μm from the focal plane.

Each mask consists of a very robust mask-frame in which a thin sheet of ceramics (thickness about 200 μm) is fixed by glue. In order to avoid flexing out of the focal plane (tolerance ± 60 μm), the sheets may have to be fixed both at the outer rim and at a central supporting cross. Mask frame and mask sheet need to have a small thermal expansion coefficient. Currently the combination of an invar frame and ceramics sheets is foreseen.

Beside the sheet, the mask frame contains two blank surfaces of 10 x 100 $mm^2$ each (see Figure 16), which are tilted by 15° with respect to the focal plane. They are needed for acquisition and guiding: several reference stars have to lie on the tilted blank surfaces. Reference marks (crosses, circles, etc.) will be cut into the surfaces by the same laser machine in the same process as cutting the slitlets.

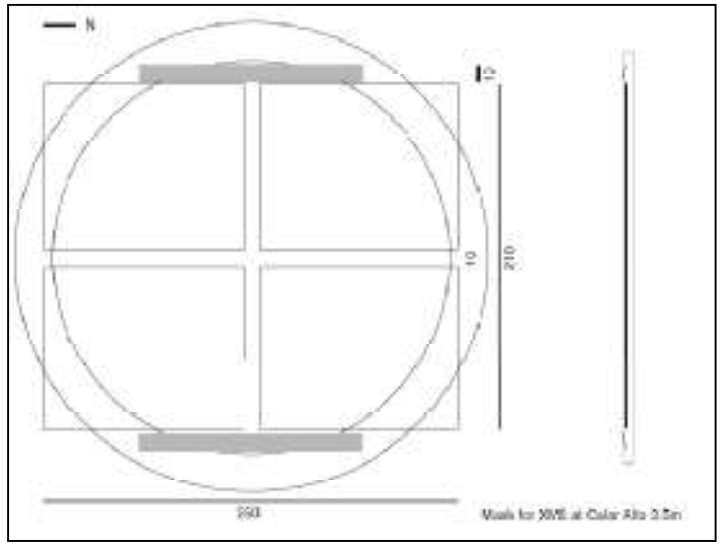

Figure 16. Top and side view sketch of a XMS mask. The tilted blank surfaces are shown.

The slitlets for the targets should have width x length = 100 μm x 650 μm (1.5" x 10") or 70 μm x 600 μm (1.0" x 9"). For some applications even shorter slits are possible. Several problems have been identified:

- Cutting 4000 slitlets in ~2 hours (i.e. cutting time = observing time) seems to require a laser cutting machine. The type of machine is still under investigation. Most likely the project needs its own cutting machine (costs 300 to 400 kEuro).

- In case the mask sheet needs to be thicker than 100μm: rectangular slit profiles (e.g. 70μm x 200μm for 1" slits in a 200μm sheet) will not act as sharp edge. Ideally the defining edge needs to be thinner than about 50μm. Thus, details how to profile the slits with a laser machine are currently investigated.

- Both the suitable ceramics material and the mask frame will be expensive (about 100 Euro per mask).

### 4.7.1. Acquisition & Guiding Unit

To secure guiding to better than +/- 0.1", it is mandatory that the guiding unit does not move by more than ± 6 μm with respect to the slit mask. This can only be achieved if guiding is implemented **on the mask** itself. So, the guide stars have to be tracked on the mask frame. On the other hand, mask acquisition needs to locate the relative position of several reference stars with respect to the slitlets on science targets within ±0.1" (~7 μm). The present concept is to reach both goals **in one** coherent fashion.

Figure 17 shows the principal layout of the acquisition and guiding unit: A small CCD camera looks onto the tilted surfaces (tilted by 15° with respect to focal plane). Two guiding units on each side of the mask are moveable along the 100 mm sides of the surfaces. In this way, **four** independent reference/guiding positions can be used. The full field on both sides together is 1/30 sqdegree. The SDSS contain typically 150 stars / sqdegree with r < 15. So 5±2 stars r < 15 should be available on average.

The 15° tilt of the surfaces will lead to images of the reference stars that are not perfectly focused near the edge of the reference fields (image size up to 4 arcsec). It has to be tested in the laboratory how much tilt/defocusing can be allowed. Alternatively, optical solutions in which the acquisition camera looks straight onto the mask (using semi-reflective mirrors) have to be investigated.

In the computer-controlled acquisition procedure two or more reference stars will be centred onto their reference position using two degrees of freedom:

1. A fine rotation of the mask by +/- 2 degrees (Figure 15 (right))to allow for misalignment of the N-S orientation with the instrument axis (limited accuracy of mounting the instrument, image rotation due to differential refraction).
2. Translation in X-Y (or Alpha-Delta) by offsetting the telescope.

After the reference stars are centred, the best of them (brightness, focus) will be used for guiding during the spectroscopic exposure. To this end the offset between star and reference position has to be monitored constantly and used to adjust the telescope tracking. Figure15 (left) shows the mask exchange mechanism with the four A&G cameras which can be moved on X-Y stages (in/out of the beam to allow mask exchange and along the tilted surfaces to find the reference star). The beam-folding prisms are replaced by flat mirrors.

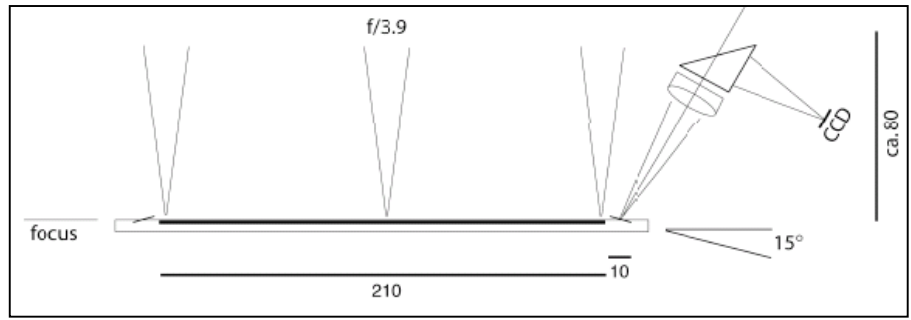

Figure 17. Schematic layout of the acquisition and guiding cameras. A total of four cameras (two for each reference surface) will be implemented.

## 5. MASS ESTIMATE

The current mass estimation (5% margin included) is about 251kg. The OB (Al/Si alloy) takes 46kg, while the Cryotiger-based Detector Dewar takes 47kg (Detectors Mount included). Noticeable is the fact that the optomechanics of the 4 CoMs take only 8.8kg (no lenses included), while all the optics are about 44kg. The MEU (steel-made) is the heaviest XMS subsystem, the main reason being that it is highly submitted mechanically since it interfaces both to the K3 Corrector and the OB. The masses for the CaMS and the CFU have been estimated according to the CoMS mass.

There is still some margin for further lightweights to be done in the next project phase. So, the safety margin may be increased. Anyway, already the present design copes with this and the other stringent requirements.

## REFERENCES

[1] Content, R., Shanks, T., "Extreme multiplex spectroscopy at wide-field 4-m telescopes", Proc. SPIE 7014, 701475 (2008)

[2] Siegel, R. and Howell, J., [Thermal radiation heat transfer], Taylor&Francis, New York, 207-250 (4$^{th}$ ed., 2002)

[3] Modest, M. F., [Radiative heat transfer], McGraw-Hill, Massachussets, chapters 4 and 5, (2$^{nd}$ ed., 2003)

[4] http://www.me.utexas.edu/~howell/tablecon.html